\documentclass[sigconf]{acmart}

\AtBeginDocument{%
  }

\copyrightyear{2026}
\acmYear{2026}
\setcopyright{cc}
\setcctype{by}
\acmConference[CHI '26]{Proceedings of the 2026 CHI Conference on Human Factors in Computing Systems}{April 13--17, 2026}{Barcelona, Spain}
\acmBooktitle{Proceedings of the 2026 CHI Conference on Human Factors in Computing Systems (CHI '26), April 13--17, 2026, Barcelona, Spain}
\acmPrice{}
\acmDOI{10.1145/3772318.3791929}
\acmISBN{979-8-4007-2278-3/2026/04}

\usepackage{xspace}
\newcommand{\oursystem}{MoXaRt\xspace}

\newcommand{\etal}{et al.\xspace}

\newcommand{\eg}{\emph{e.g.,}\xspace}
\newcommand{\dq}[1]{``#1''\xspace}
\newcommand{\sq}[1]{`#1'\xspace}

\usepackage{booktabs}
\usepackage{pifont} 
\usepackage{graphicx}
\usepackage{multirow}

\newcommand{\cmark}{\ding{51}}%
\newcommand{\xmark}{\ding{55}}%

\usepackage{enumitem}
\usepackage{subfigure}
\usepackage{listings}

\usepackage{url}
\usepackage{fancyvrb}

\definecolor{codegreen}{rgb}{0,0.6,0}
\definecolor{codegray}{rgb}{0.5,0.5,0.5}
\definecolor{codepurple}{rgb}{0.58,0,0.82}
\definecolor{backcolour}{rgb}{0.95,0.95,0.92} 
\lstdefinestyle{mystyle}{
    backgroundcolor=\color{backcolour},
    commentstyle=\color{codegreen},
    keywordstyle=\color{magenta}, 
    numberstyle=\tiny\color{codegray},
    stringstyle=\color{codepurple},
    basicstyle=\ttfamily\normalsize, 
    breakatwhitespace=false,
    breaklines=true,                 
    captionpos=b,                    
    keepspaces=true,
    numbers=left,                    
    numbersep=5pt,
    showspaces=false,
    showstringspaces=false,
    showtabs=false,
    tabsize=2,
    frame=single,                    
    rulecolor=\color{black},         
    title=\lstname                   
}
\newif \ifdraft \drafttrue   

\newif \ifhighlight \highlighttrue    
\newif \ifhighlight \highlightfalse    

\newcommand{\change}[1]{#1}

\begin{document}

\title{\oursystem: Audio-Visual Object-Guided Sound Interaction for XR}


\settopmatter{authorsperrow=3}

\author{Tianyu Xu}
\orcid{0009-0009-9135-6080}
\affiliation{%
  \institution{Google}
  \city{Mountain View}
  \state{CA}
  \country{USA}
}
\email{tyx@google.com}

\author{Sieun Kim}
\affiliation{%
  \institution{University of Michigan}
  \city{Ann Arbor}
  \state{MI}
  \country{USA}
}
\email{sieunk@umich.edu}

\author{Qianhui Zheng}
\affiliation{%
  \institution{University of Michigan}
  \city{Ann Arbor}
  \state{MI}
  \country{USA}
}
\email{qianhui@umich.edu}

\author{Ruoyu Xu}
\affiliation{%
  \institution{Columbia University}
  \city{New York}
  \state{NY}
  \country{USA}
}
\email{rx2241@columbia.edu}

\author{Tejasvi Ravi}
\affiliation{%
  \institution{Google}
  \city{San Francisco}
  \state{CA}
  \country{USA}
}
\email{ravitejasvi@google.com}

\author{Anuva Kulkarni}
\affiliation{%
  \institution{Google}
  \city{Mountain View}
  \state{CA}
  \country{USA}
}
\email{anuvak@google.com}

\author{Katrina Passarella-Ward}
\affiliation{%
  \institution{Google}
  \city{San Francisco}
  \state{CA}
  \country{USA}
}
\email{kpassarella@google.com}

\author{Junyi Zhu}
\authornotemark[1]
\affiliation{%
  \institution{University of Michigan}
  \city{Ann Arbor}
  \state{MI}
  \country{USA}
}
\email{zhujunyi@umich.edu}

\author{Adarsh Kowdle}
\authornote{Co-last authors, equal contribution.}
\affiliation{%
  \institution{Google}
  \city{San Francisco}
  \state{CA}
  \country{USA}
}
\email{adarshkowdle@google.com}

\renewcommand{\shortauthors}{Xu et al.}

\begin{abstract}
In Extended Reality (XR), complex acoustic environments often overwhelm users, compromising both scene awareness and social engagement due to entangled sound sources. We introduce \oursystem, a real-time XR system that uses audio-visual cues to separate these sources and enable fine-grained sound interaction. \oursystem's core is a cascaded architecture that performs coarse, audio-only separation in parallel with visual detection of sources (\eg faces, instruments). These visual anchors then guide refinement networks to isolate individual sources, separating complex mixes of up to 5 concurrent sources (\eg 2 voices + 3 instruments) with $\sim$2 second processing latency. We validate \oursystem through a technical evaluation on a new dataset of 30 one-minute recordings featuring concurrent speech and music, and a 22-participant user study. Empirical results indicate that our system significantly enhances speech intelligibility, yielding a 36.2\% ($p < 0.01$) increase in listening comprehension within adversarial acoustic environments while substantially reducing cognitive load ($p < 0.001$), thereby paving the way for more perceptive and socially adept XR experiences.
\end{abstract}


\begin{teaserfigure}
\vspace{5pt}
  \includegraphics[width=\textwidth]{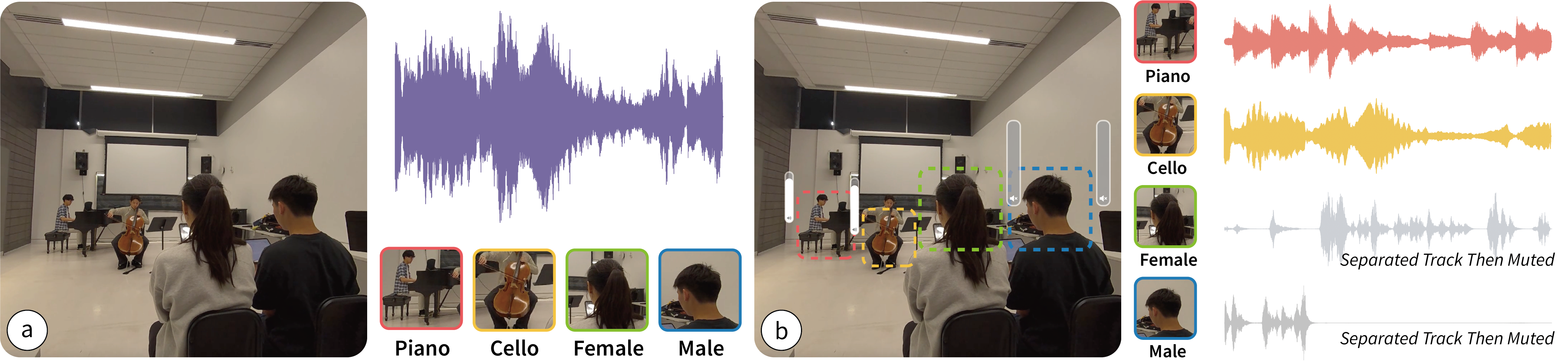}
  \caption{\oursystem is a real-time XR system that (a) uses audio-visual cues to separate sound sources from a single audio channel and (b) enables fine-grained, interactive control over the user’s soundscape.}
\vspace{5pt}
  \label{fig:teaser}
\end{teaserfigure}

\maketitle

\section{Introduction}
\label{sec:introduction}


Modern Extended Reality (XR) glasses and headsets grant users unprecedented control to filter, enhance, and overlay digital information onto their visual world \cite{KinectFusion, holoportation, GesturAR}. Yet, this level of fine-grained, interactive control over perception has not extended to the auditory realm \cite{Moustakas, Yang2022Audio}. In complex acoustic environments, XR users are often overwhelmed by a cacophony of sounds, making it difficult to focus on what truly matters \cite{sweller1988cognitive}. This challenge stems from the user’s fundamental inability to disentangle multiple simultaneous sound sources \cite{cocktail-party-1953}. Unlike visual elements, which can be digitally manipulated, the ambient audio of people talking, instruments playing, and background noise remains a largely unfiltered and entangled mixture \cite{Yang2022Audio}. The result is compromised scene awareness and social engagement; important sounds, such as a specific conversation or a lead musician's solo, are drowned out, which impairs comprehension and increases cognitive load \cite{sweller1988cognitive}.

The core difficulty lies in performing real-time sound source separation on a head mounted device. The acoustic scene in XR is highly non-stationary, as head and sound source motion continually alter directions of arrival and room acoustics~\cite{audio-scene-complexity-2021}. Furthermore, microphones on head-worn devices provide a limited spatial aperture due to their compact form factor, resulting in effectively mono or near-mono auditory inputs despite multiple sensors when sound sources are at distant (\eg > 3m)~\cite{sounddistance2022}. This hardware constraint, combined with the complexity of real-world auditory content—where human speech, music, and ambient noise constantly overlap and mask one another—makes robust audio separation based solely on a headset's microphone inputs an extremely challenging problem. Past approaches have either relied on cumbersome hardware setups \eg large microphone arrays unsuitable for wearable use~\cite{microphone-array-2022}, or have been computationally too expensive for real-time interaction~\cite{crab-cvpr2025, DAVIS}.

We introduce \oursystem, a real-time XR system that tackles this challenge by leveraging audio-visual cues to separate sound sources from a single audio channel and enable fine-grained, interactive control over the user's soundscape. \oursystem's core is a cascaded architecture operating in parallel: it performs a coarse, audio-only separation to split the soundscape into general categories (\eg speech, music, and noise) while concurrently running face and instrument detection to identify visual anchors in the video stream. These anchors then guide specialized refinement networks, which isolate individual speakers from the main speech track and specific instruments from the music track. This process generates a live audio-visual map of the environment, identifying \dq{who or what is making each sound.} Based on this map, \oursystem empowers users with a personal \dq{audio mixer} for the world, allowing them to isolate a single speaker in a noisy room, adjust the volume of individual instruments in a band, or remix their entire acoustic environment in real time.

We validate \oursystem through both technical evaluation and a user study. For our technical evaluation, we collected a new, challenging real-world audio-visual dataset featuring complex mixtures of up to five concurrent speakers and three musical instruments. On this dataset, \oursystem's sound separation performance surpasses state-of-the-art audio-visual models. Our cascaded architecture successfully separates up to four distinct human voices or a mix of five total sources in real time, with a consistent processing latency of around 2 seconds. Finally, a 22-participant user study demonstrated \oursystem's practical benefits across six real-world scenarios, from a bustling concert to a multi-group meeting. With \oursystem enabled, participants performed significantly better at understanding target speech in noisy conditions, achieving an average improvement of 36.2\% ($p=0.0058$) in a multi-speaker listening comprehension task, and they reported significantly higher clarity and \change{lower cognitive load}. 


In summary, we contribute:

\begin{itemize}
\item \oursystem, the first XR system to integrate real-time, audio-visual sound separation as a core primitive for direct, interactive control over a user's authentic soundscape.
\item A cascaded audio-visual transformer model that leverages visual cues to perform robust, real-time separation of multiple speech and non-speech sources from a single-channel audio input.
\item A real-world audio-visual dataset featuring complex scenes with multiple concurrent speakers and musical instruments, designed for evaluating interactive sound separation systems.
\item A comprehensive evaluation demonstrating our model's state-of-the-art technical performance and a user study confirming that \oursystem significantly improves communication clarity, \change{reduces cognitive load}, and enhances user experience in complex auditory environments.
\end{itemize}

\section{Related Work}
\label{sec:related_work}

\subsection{Sound Interaction in XR}
\label{sec:sound_interaction_in_xr}

Audio is a crucial component for creating persuasive and immersive experiences in XR. Foundational research has centered on high-fidelity spatialization, using technologies like binaural rendering and Head-Related Transfer Functions (HRTFs) to create a sense of presence~\cite{spatial_audio_in_vr, sound_vr}. Recent work continues to enhance this immersion by leveraging multimodal scene understanding to create context-aware auditory experiences~\cite{Auptimize2024, SonifyAR_UIST24, Samosa2025, nongpiur2025system, nongpiur2025passive}, with industry leaders investing heavily in making virtual sounds perceptually indistinguishable from reality~\cite{meta2022audiopresence}. However, this blending of real and virtual soundscapes in Mixed Reality (MR) introduces new challenges. The accumulation of sounds from users' distinct physical environments often leads to acoustic incoherence and increased cognitive load~\cite{audio_social_interacton, SoundShift-dis2024}. \change{This necessitates a paradigm shift: moving from simply rendering sound to enabling users to manage and manipulate their auditory environment, a goal Haas \etal formalized as Interactive Auditory Mediated Reality (IAMR)~\cite{Haas2020IAMR}.}

In response to this need, the HCI community has developed compelling paradigms that treat audio as an active medium. A major thread focuses on augmenting or filtering the soundscape at a global or class level. For instance, Wu \etal's \textit{New Ears} allows users to spatially decouple their listening point from their physical location to facilitate object search~\cite{NewEars-ismar2024}, while Chang \etal's \textit{SoundShift} manages information overload by spatially shifting notifications or altering the transparency of real-world audio~\cite{SoundShift-dis2024}. \change{Similarly, semantic filtering systems like \textit{Semantic Hearing}~\cite{SemanticHearing} and \textit{ProtoSound}~\cite{ProtoSound2022} enable users to suppress or prioritize specific sound categories (\eg vacuums, speech) or personalize sound events for accessibility.}

More recently, research has shifted toward object-centric interaction. \textit{SonoHaptics} maps visual properties to audio-haptic feedback to aid cursor-less gaze-based selection~\cite{SonoHaptics-uist2024}, while \textit{AudioMiXR} treats virtual audio as a tangible object for 6DoF spatial manipulation~\cite{AudioMiXR_IMWUT25}. Similarly, \textit{SonifyAR} and \textit{Sonify Anything} use visual recognition to synthesize context-aware sound effects~\cite{SonifyAR_UIST24, SonifyAnything_ISMAR25}. \change{However, these systems rely on generative or synthetic audio rather than processing the object's actual acoustic emission. In the domain of veridical audio, \textit{Look Once to Hear} enables users to isolate a specific speaker by looking at them~\cite{LookHear-chi2024}. Yet, this approach requires an explicit enrollment phase and functions as a binary filter (target vs. noise), lacking the ability to dynamically remix multiple concurrent sources (\eg music and speech). The critical limitation remains: no existing XR system allows users to seamlessly separate, balance, and remix the live, veridical sounds of multiple real-world objects in real-time.}

\subsection{Sound Separation in Machine Learning}
\label{sec:sound_separation_in_ml}
Parallel to these HCI advancements, the signal processing and machine learning communities have developed sophisticated technologies for analyzing and separating sound sources. The classical cocktail party problem, which involves isolating individual speakers from a mixture of voices~\cite{cocktail-party-1953}, has evolved from early signal processing methods like Independent Component Analysis (ICA)~\cite{ica_2000} and blind source separation (BSS)~\cite{bss_survey} to modern deep learning approaches that achieve remarkable separation quality~\cite{deep-survey-2023}.

\subsubsection{Audio-Only and Universal Separation}
Modern audio-only source separation is characterized by Transformer-based architectures, with models like MossFormer2~\cite{mossformer2} achieving state-of-the-art results on speech separation benchmarks. \change{In the music domain, specialized architectures like the Band Split Roformer (BSR) have achieved high-fidelity isolation of individual instruments~\cite{Music_BSR_2024, Demucs_2023}.} The field has also expanded to address the \dq{cocktail fork problem}~\cite{cocktail-fork-2022}, separating complex soundtracks into speech, music, and sound effects simultaneously, while other work leverages spatial cues and linguistic context for simultaneous translation~\cite{SpatialTranslation-chi2025}. Universal sound separation systems~\cite{universal-ss-2019} attempt to separate arbitrary sources without prior knowledge, though performance varies across categories. \change{Building on these approaches, Target Sound Extraction (TSE) focuses on isolating specific sources based on text labels or reference audio~\cite{RealTimeTSE_Veluri23, SoloAudio2025}. While recent work has explored real-time TSE~\cite{RealTimeTSE_Veluri23} and wearable separation systems like \textit{ClearBuds} and \textit{Look Once to Hear}~\cite{ClearBuds_2022, LookHear-chi2024}, these typically function as binary extractors or speech enhancers, lacking the visual grounding and continuous, multi-source mixing control required for intuitive XR interaction.}

\subsubsection{Audio-Visual Separation}
A significant breakthrough has been the incorporation of visual information to guide audio separation. \change{This domain was pioneered by \textit{The Sound of Pixels}~\cite{SoundOfPixels2018} and \textit{Looking to Listen}~\cite{LookingToListen2018}, which demonstrated that vision could act as a precise filter for both musical instruments and human speech. Building on this, \textit{VisualVoice} introduced cross-modal consistency to leverage facial appearance as a strong prior for separating speech~\cite{VisualVoice_2021}.} Modern architectures extend these principles: \textit{AudioScopeV2}~\cite{audioscopev2} uses cross-modal attention to bind on-screen objects to sounds, \change{while \textit{iQuery} formulates separation as a query-based task, utilizing detected visual objects to initialize learnable audio queries~\cite{iQuery_CVPR23}.} More recent work, like \textit{Crab}~\cite{crab-cvpr2025} and \textit{AVMossformer2}~\cite{clearervoice}, employs transformer architectures to learn joint representations. Furthermore, advanced generative approaches, such as \textit{DAVIS}~\cite{DAVIS}, employ diffusion models to synthesize separated sound.

Despite these advances, current ML approaches face significant limitations when deployed for interactive applications. Most models are designed for offline processing, requiring the entire audio sequence for optimal performance~\cite{SoundOfPixels2018, LookingToListen2018, audioscopev2, crab-cvpr2025, DAVIS}. Real-time operation, a prerequisite for interactive XR, often necessitates compromises in separation quality or introduces unacceptable latency and computational requirements that exceed the capabilities of consumer XR devices, particularly in generative diffusion models~\cite{DAVIS, SoloAudio2025}. Furthermore, these systems typically treat separation as a passive signal processing task without human input in the loop. \change{For example, they can isolate a violin from an orchestra in a recording, but they cannot respond to a user pointing at a specific violin in a live performance and saying \dq{make that one louder} while preserving the veridical sound of that instrument.}


\change{The preceding review highlights a critical disconnect between the high-level interaction needs of XR soundscapes and the low-level capabilities of current sound separation models. While HCI research manipulates live audio through global filters or generative synthesis, it lacks frameworks for object-centric, veridical interaction with live signals. Conversely, ML models isolate authentic sources but typically function as passive, offline tasks. Consequently, no prior work has operationalized visually-guided source separation as a real-time primitive for direct user interaction in XR.}


\section{\oursystem}
\label{sec:system}

\begin{figure*}[h]
\includegraphics[width=2\columnwidth]{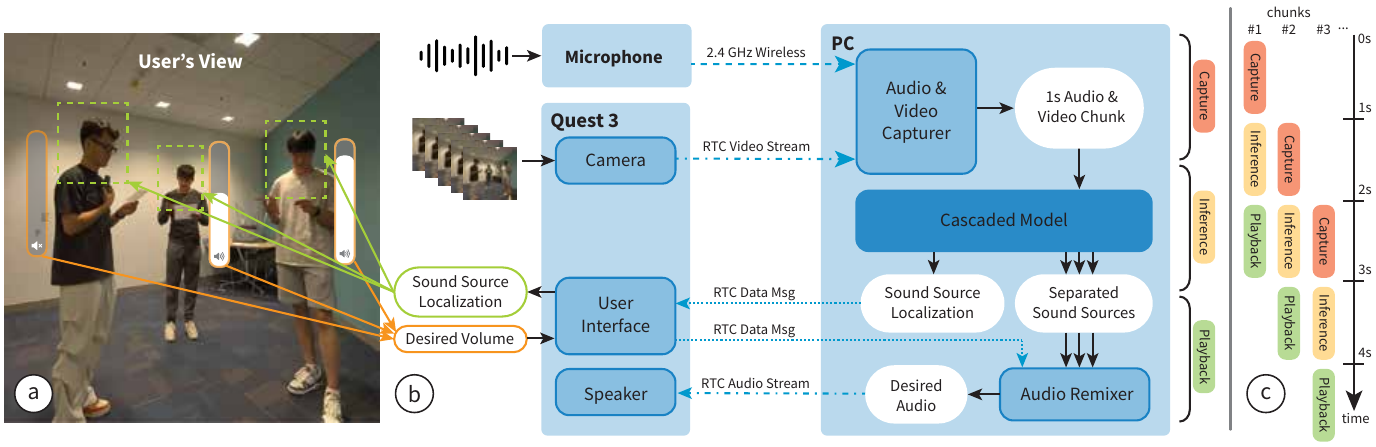}
\caption{The user interface of \oursystem and walkthrough of its internal implementation. (a) The user can identify separated sound sources through visualizations and adjust the volume of each source for remixing. (b) Video captured from the Quest 3 headset is streamed to the PC for model inference, and the remixed audio is streamed back. (c) To ensure continuous playback of remixed audio, \oursystem operates in a three-stage pipeline: Video and Audio Capture, Model Inference, and Playback.}
\label{fig:app_walkthrough}
\end{figure*}

To enable intuitive control over complex acoustic scenes, we present \oursystem, an XR system utilizing a cascaded, coarse-to-fine architecture for interactive audio-visual source separation. The user interface (Sec.~\ref{sec:user_interface}) visualizes separated sources and allows users to adjust individual volumes for real-time audio remixing (Sec.~\ref{sec:remixing}). By offloading model inference to an external PC via wireless transmission (Fig.~\ref{fig:app_walkthrough}), the system ingests multi-modal streams through a parallel analysis stage (Sec.~\ref{sec:coarse_separation}). The resulting outputs serve as contextual priors for specialized Music Refinement (Sec.~\ref{sec:music_refinement}) and Speech Refinement (Sec.~\ref{sec:speech_refinement}) networks. The following sections detail \oursystem's user interface and model architecture.

\subsection{User Interface}
\label{sec:user_interface}
\subsubsection{User Interaction}

Our headset-based XR interface visualizes localized sound sources, allowing users to control individual volumes. Each model-detected source is represented by a visual marker at its spatial position, enabling rapid identification of multiple concurrent sounds. Users can select specific sources and adjust their levels via interactive sliders (operated through controllers or hand gestures) to customize audio mixing for varying scenarios. To ensure an immersive experience, the remixed audio is delivered via noise-canceling earbuds connected to the XR headset, effectively isolating the user from ambient interference. 


\subsubsection{Implementation}
\label{sec:interface_implementation} 

We implemented \oursystem for the Meta Quest 3 headset\footnote{Meta Quest 3. \url{https://www.meta.com/quest/quest-3/}} using Unity 6000.2.2f1\footnote{Unity. \url{https://unity.com}}. Model inference is offloaded to a dedicated PC equipped with an NVIDIA RTX 5080 GPU. To ensure high-quality audio capture, we use a Rode Wireless GO II\footnote{Rode Wireless GO II. \url{https://rode.com/en-us/products/wirelessgoii}} microphone, which transmits wirelessly to a receiver connected to the PC. The system utilizes WebRTC for bidirectional, real-time data transmission: the headset streams captured video to the PC, while the processed audio is returned to the headset. This remixed audio is played through active noise-canceling earbuds to eliminate acoustic interference between the processed output and the physical environment. We opted for an external microphone because Meta's audio API restricts access to raw signals, providing only heavily filtered audio. To maintain consistency with typical headset hardware, the external microphone is set to mono-mode for all applications and user study scenarios presented in this work.

\paragraph{Processing Pipeline}

To support real-time sound separation and remixing, our interface operates in a three-stage pipeline: \textit{Capture} (video and audio), \textit{Inference} (model processing), and \textit{Playback} (Fig.~\ref{fig:app_walkthrough} (c)). Each stage processes data in 1-second segments, resulting in continuous playback with a constant latency of 2 seconds. During the \textit{Capture} stage, camera frames from the Quest 3 and audio from the Rode microphone are synchronized into 1-second chunks. These chunks are appended to a 1-minute rolling buffer, which provides the necessary temporal context for our model to ensure consistent source separation. In the \textit{Inference} stage, this 1-minute window serves as input to our cascaded model. The model identifies all speakers and musical instruments, separating them into individual tracks and providing spatial coordinates for localization. Only the final 1 second of the separated audio, corresponding to the most recent input chunk, is retained for remixing. Finally, in the \textit{Playback} stage, the audio remixer applies user-defined volume levels to each source and merges them into a single output. This remixed stream is transmitted via WebRTC and played through the user's earbuds.
\paragraph{Sound Source Localization}

Upon deployment in a new environment, \oursystem processes the initial video stream through the cascaded model to identify and localize all individual sound sources. The model generates coordinates for each source within the video frames, which are transmitted to the headset via RTC data messages and visualized as markers in the interface (shown in green in Fig.~\ref{fig:app_walkthrough} (a)). Users can then adjust the volume of specific sources using interactive bars (shown in orange in Fig.~\ref{fig:app_walkthrough} (a)) via controllers or hand gestures. These updated volume levels are sent back to the PC via RTC data messages. This feedback loop ensures that separated audio from subsequent inference cycles is remixed according to user preferences before being returned to the headset through the established RTC audio stream.

\paragraph{Audio Remixing}
\label{sec:remixing}

The audio remixer module synthesizes the final auditory scene by combining the separated audio stems with the user-defined gain levels. Let the set of $N$ refined audio stems from the preceding stages be $\{s'_1, s'_2, \dots, s'_N\}$. Based on user input via the interface, a corresponding gain vector $g = [g_1, g_2, \dots, g_N]$ is determined, where each $g_n \in [0, 10]$ represents the desired amplification factor for its respective source. The final monophonic audio output $y_{\text{final}}$ is synthesized as a weighted sum:

\[y_{\text{final}} = \sum_{n=1}^{N} g_n s'_n\]
This composite signal is subsequently rendered into a binaural stereo stream to provide the user with a spatially coherent auditory experience.

\subsection{Model}

\begin{figure*}[h]
\includegraphics[width=1.7\columnwidth]{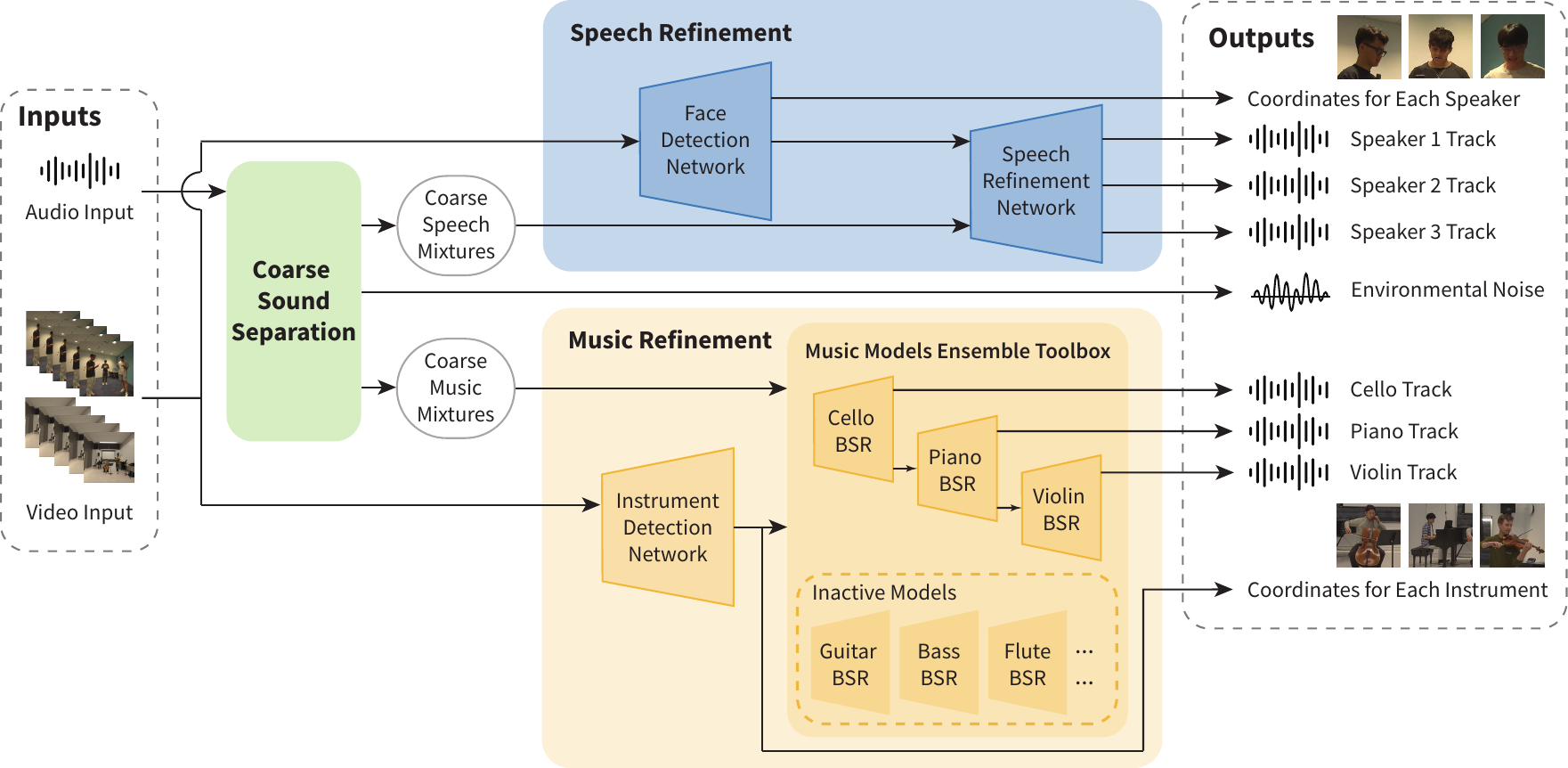}
\caption{The cascaded architecture of \oursystem for multi-modal sound separation. The system pipeline consists of two main stages: First, a \textbf{Coarse Sound Separation} module processes the raw audio inputs to produce initial coarse speech and music mixtures. Next, these mixtures, along with outputs from the \textbf{Face Detection Network} and \textbf{Instrument Detection Network}, are fed into two parallel refinement branches: \textbf{Speech Refinement} and \textbf{Music Refinement}. The Speech Refinement branch further disambiguates the coarse speech into individual speaker tracks and their corresponding spatial coordinates. The Music Refinement branch utilizes an ensemble of \textbf{Band Split Roformer (BSR)} models, dynamically activated by the detected instruments, to output isolated instrument tracks. This structured approach enables precise source separation by leveraging visual context only when specialized refinement is required.}
\label{fig:pipeline}
\end{figure*}

\subsubsection{Coarse Sound Separation}
\label{sec:coarse_separation}
The Coarse Sound Separation module provides an initial estimate of the constituent sources within the input audio. For this task, we adapt the separation module $\mathcal{M}^S$ from AudioScopeV2 \cite{audioscopev2} that is based on the dilated convolutional network with a learnable encoder-decoder architecture. The network, defined by the function $f_{\theta}$, processes a raw mixed waveform $x \in \mathbb{R}^T$ to estimate $\mathcal{M}=3$ source waveforms $\{\hat{s}_m\}_{m=1}^{\mathcal{M}}$, where $\hat{s}_m \in \mathbb{R}^T$ corresponding to speech, music, and environmental noise. The separation is performed by applying the model to the mixture:
\[\{\hat{s}_1, \hat{s}_2, \hat{s}_3\} = f_{\theta}(x)\]
Following the original training protocol, we incorporate a mixture consistency layer to enforce the constraint that the separated outputs sum to the original input:
\[x = \sum_{m=1}^{\mathcal{M}} \hat{s}_m\]
In contrast to prior audio-visual methods \cite{audioscope, LookingToListen2018, DAVIS}, we adopt a purely audio-based approach for this initial stage. This design is motivated by two factors. First, it sidesteps the significant computational overhead of visual conditioning, which often provides only marginal gains in coarse separation. Furthermore, this modularity avoids the rigidity of monolithic end-to-end architectures; our pipeline enables dynamic resource allocation by activating specialized refinement modules only when their specific visual anchors are detected. Second, and more importantly for our cascaded design, it allows us to pre-train separation network on large-scale dataset corpora (\eg YFCC100M \cite{YFCC100M}) using self-supervised techniques like MixIT \cite{wisdom_2020}. This robust, audio-only pre-training provides a generalized and high-quality initial separation, which is essential for the downstream Speech and Music Refinement modules to work effectively as illustrated in Fig.~\ref{fig:pipeline}.

\subsubsection{Music Refinement}
\label{sec:music_refinement}
To give users fine-grained control over musical elements, such as isolating an instrument for learning or creating a live karaoke mix, the Music Refinement module transforms the coarse music estimate into high-fidelity, separated instrument stems. It first employs a visual Instrument Detection network to identify which instruments are present. This visual context then dynamically activates specific models within our high-fidelity teacher model, whose knowledge is later distilled into an efficient student model.


\paragraph{Instrument Detection}
\label{sec:instrument_detection}
First, a visual Instrument Detection network identifies which of $K=7$ target instruments are present. For this, we use a multi-head DeepLabv3+ architecture \cite{chen_2018}, with a MobileNetV2 backbone \cite{MobileNetV2}, which is a standard choice for efficient on-device semantic segmentation. The backbone extracts multi-scale features which are processed by an Atrous Spatial Pyramid Pooling (ASPP) decoder. Following the DeepLabv3+ design, low-level features are fused with the ASPP output before being passed to a dedicated instrument classification head, which is trained to predict the presence of 7 relevant instrument classes: Vocal, Piano, Violin, Cello, Bass, Guitar, and Flute.

To train this model, we curated a large-scale dataset using an automated labeling pipeline. Starting with the COCO-Stuff dataset \cite{cocostuff}, we first employed the Segment Anything Model (SAM) \cite{SAM} to generate high-quality instance masks for all objects in each image. Following the method of Peng et al. \cite{peng_2023}, a Vision-Language Model (VLM) then automatically assigned an instrument label to each mask by comparing the semantic similarity of the cropped image region with text embeddings for our target classes. We trained the instrument classification head using a transfer learning strategy, freezing the pre-trained MobileNetV2 backbone and fine-tuning only the parameters of the new head and the ASPP decoder. We treat this as a multi-label classification problem and train the network using a Binary Cross-Entropy (BCE) loss with logits across the $K$ instrument classes.

\paragraph{Dynamic Separation Ensemble (Teacher)}
\label{sec:music_ensemble_teacher}
Our high-fidelity teacher model for music is a dynamic ensemble of specialist networks. This module leverages predictions from the Instrument Detection network to perform targeted, high-quality separation. The ensemble consists of seven specialist models, one for each target instrument, based on the state-of-the-art Band Split Roformer (BSR) architecture \cite{Music_BSR_2024}. To optimize efficiency, the system utilizes dynamic activation where only the BSR models corresponding to instruments detected with a confidence above a threshold ($\tau = 0.5$) are executed. This sparse and adaptive approach significantly reduces computational overhead while maintaining the high fidelity required to serve as an ideal data generator for our student model.

\paragraph{Distilled Real-Time Model (Student)}
\label{sec:music_ensemble_student}
Although the teacher ensemble provides state-of-the-art fidelity, its computational footprint is prohibitive for real-time XR applications. To address this, we employ model distillation \cite{hinton2015distill}. Let the teacher be represented by a function $T$ and the lightweight student model be $S_{\phi}$ with parameters $\phi$. We generate a dataset of pairs consisting of the coarse track $x_c$ and pseudo-target stems $\hat{y}_T$, where $\hat{y}_T = T(x_c)$. The student is trained to mimic the teacher by optimizing its parameters $\phi$ to minimize a signal-level loss $\mathcal{L}_{\text{signal}}$:$$\min_{\phi} \mathcal{L}_{\text{signal}}(S_{\phi}(x_c), \hat{y}_T)$$For $\mathcal{L}_{\text{signal}}$, we utilize the negative scale-invariant signal-to-distortion ratio (SI-SDR) loss \cite{sisdr}. This strategy effectively transfers the capabilities of the specialist ensemble into a single efficient model.

\subsubsection{Speech Refinement}
\label{sec:speech_refinement}
Concurrent with music refinement, this module resolves the coarse speech signal into discrete stems for each speaker. The system utilizes the video feed to disambiguate speakers based on their facial information, allowing the audio to be mapped to specific individuals in the scene.

\paragraph{Face Detection}
\label{sec:face_detection}
To associate speech with a specific speaker, we adapt the yolov8-face model \cite{yolov8face}. This open-source project fine-tunes the YOLOv8 framework \cite{ultralytics} specifically for face detection tasks. The model provides real-time bounding boxes that serve as the visual anchors required to guide the downstream separation process.

\paragraph{Online and Offline Separation Models}
\label{sec:speech_separation_models}
Our system supports two distinct processing modes. For offline processing, we employ AV-MossFormer2 \cite{clearervoice} as a high-fidelity teacher model, $T_{AV}$. For online processing, we distill the teacher's knowledge into a lightweight student model, $S_{\psi}$, which is based on the AudioScopeV2 \cite{audioscopev2} architecture. Given a coarse speech track $x_{sp}$ and the corresponding visual face crops $v$, the student is trained on the teacher outputs $\hat{y}_{T_{AV}} = T_{AV}(x_{sp}, v)$. The student parameters $\psi$ are optimized by minimizing the same distillation loss described in Sec.~\ref{sec:music_ensemble_student}.

\subsubsection{Implementation}
\label{sec:implementation}
We deployed and evaluated \oursystem using the hardware setup described in Sec.~\ref{sec:interface_implementation}. The visual frontend consists of the face detection model (3.2M parameters, 12.1~MB) and the instrument detection model (33M parameters, 132~MB). The audio pipeline begins with the Coarse Sound Separation module (6M parameters, 24~MB), followed by the distilled online student models for Speech Refinement (9M parameters, 36~MB) and Music Refinement (7M parameters, 28~MB). In the current configuration using these student models, the end-to-end processing latency is approximately 2~s. While these models are designed to operate within an interactive budget, this latency figure serves as a functional baseline for the cascaded architecture. We provide a comprehensive breakdown of latency components and discuss paths toward sub-second real-time performance in Sec.~\ref{sec:real-time-analysis}.
\section{Application Scenarios}
\label{sec:applications}

The ability of \oursystem to dissect a complex acoustic scene into its constituent parts and grant the user interactive control over them unlocks a wide range of applications. We highlight three scenarios that demonstrate the system's potential to reshape how we experience and interact with our auditory environment.

\begin{figure}[h]
\includegraphics[width=1.0\columnwidth]{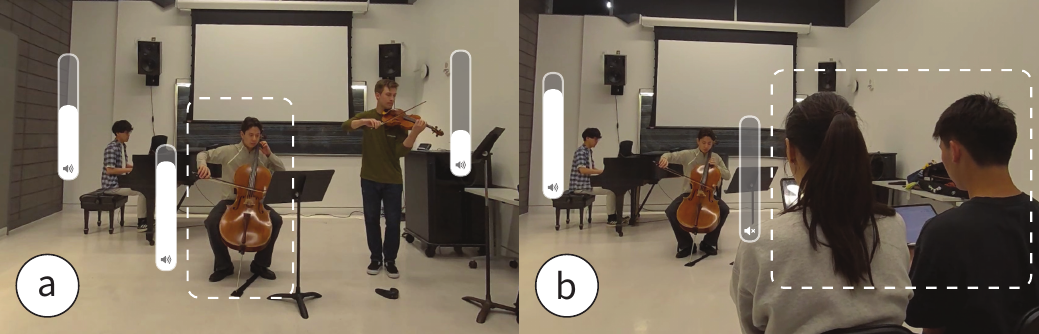}
\caption{Interactive Music Experiences. \oursystem enables users to actively control their listening experience during live music events. (a) The user can act as a real-time audio engineer, selectively adjusting the gain of individual instruments, such as the violin, cello, and piano in a trio—to create a personalized mix. (b) In a concert setting, the user can enhance acoustic focus by isolating the musicians' performance from distracting audience chatter.}
\label{fig:app-1}
\end{figure}

\subsection{Interactive Music Experiences}

In settings like live concerts or music rehearsals, \oursystem transforms the user from a passive listener into an active participant in the audio experience (Fig. 4). A user attending a concert can act as their own real-time sound engineer, adjusting the audio mix to their personal preference—for instance, boosting the violin during a solo or lowering the piano to focus on the cello's bassline. This capability is also valuable for music education, as a student could isolate a specific instrument within an ensemble to better study its part. Furthermore, the system enhances listening clarity by functioning as an intelligent de-noiser. By isolating the desired performance, a user can suppress distracting ambient sounds, such as audience chatter, to maintain an immersive and focused experience. 

\subsection{Enhanced Social Interaction}

\begin{figure}[h]
\includegraphics[width=1.0\columnwidth]{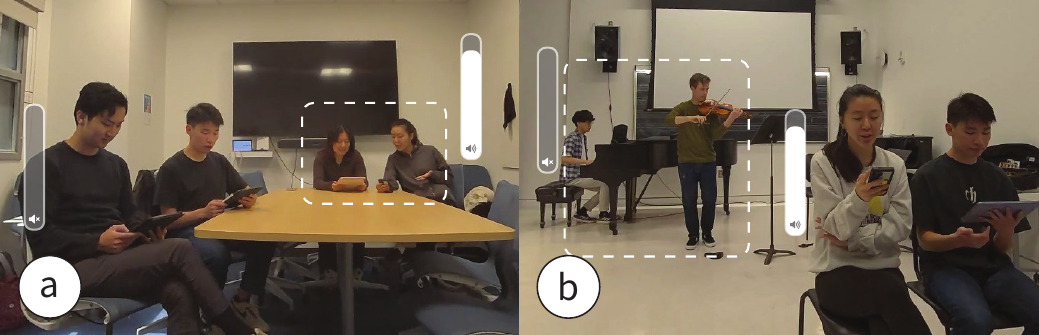}
\caption{Enhanced Social Interaction. \oursystem facilitates communication in complex and noisy social settings. (a) In an environment with multiple simultaneous conversations, a user can selectively amplify a target conversation to maintain focus. (b) The system can improve speech intelligibility in chaotic environments by isolating a conversation partner's voice from loud background noise, such as live music.}
\label{fig:app-2}
\end{figure}

\oursystem provides a powerful tool for navigating the challenges of complex social environments, effectively offering a practical solution to the classic \dq{cocktail party problem} (Fig. \ref{fig:app-2}). At a crowded reception or in a noisy café, a user can dynamically manage their auditory focus and stay engaged in multiple conversations. For example, they can choose to amplify a specific conversation, even one happening at a distance, while attenuating other nearby voices and background noise. This selective attention reduces the cognitive load associated with straining to hear in loud settings, making social engagement less fatiguing and more accessible, particularly for individuals with hearing difficulties. The user can seamlessly switch their focus between different conversational groups, giving them unprecedented control over their social soundscape.

\begin{figure}[h]
\includegraphics[width=1.0\columnwidth]{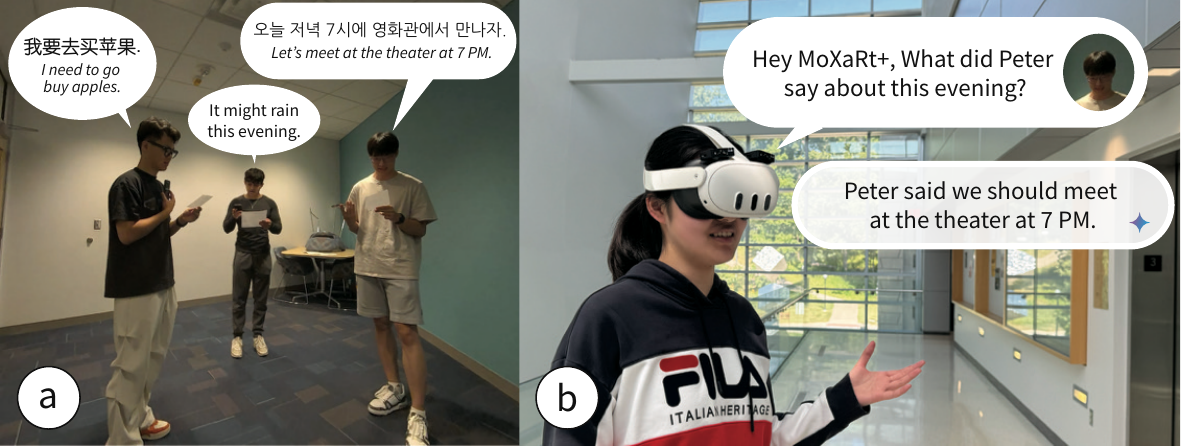}
\caption{Downstream AI Assistance. The separated streams from \oursystem can serve as clean, audio-visually aligned input for LLM assistance (\eg Gemini, Chatgpt.) (a) The system can isolate individual speakers in a multi-lingual environment, enabling real-time, multi-person transcription and translation. (b) The system's output could be used to create a queryable log of interactions, allowing a user to ask questions about past events (\eg \dq{What did Peter say about this evening?}) and receive a contextually grounded response.}
\label{fig:app-3}
\end{figure}

\change{\subsection{Future Application: Context-Aware AI Assistance}

Beyond direct human consumption, the clean, object-centric audio-visual streams produced by \oursystem serve as a critical primitive for intelligent, context-aware AI assistants\cite{google_gemini, openai_chatgpt}. Current voice-based agents typically operate on a single, mixed-audio stream, causing them to fail in \dq{cocktail party} scenarios where overlapping speech confounds Automatic Speech Recognition (ASR) engines.

By feeding \oursystem's separated audio outputs into downstream models, we can unlock two specific technical capabilities that are currently intractable in mixed-reality environments:

\subsubsection{Real-Time Diarized Translation} In a multi-lingual meeting, \oursystem can enable per-speaker translation. The system buffers the separated audio stream $s'_i$ for each detected speaker. These streams are processed in parallel by a streaming ASR model to generate text. Because \oursystem reduces the Word Error Rate (WER) by significant margins compared to raw audio (as shown in Table \ref{tab:tech_eval_results}), the ASR engine receives a signal that is actually intelligible, unlike the raw mixture. The resulting text is translated and rendered as a subtitle floating above the corresponding speaker's coordinate box (Fig.~\ref{fig:app-3}). The primary challenge here is cascading latency. The total latency $L_{total} = L_{sep} + L_{ASR} + L_{trans}$. While \oursystem’s current separation latency is $\approx 2s$, optimizing the pipeline to use sliding-window inference with smaller chunks could bring this closer to the perceptual threshold required for live captioning.

\subsubsection{Queryable RAG-based Audio-Visual Memory} Current memory in AI assistants is limited because it cannot robustly distinguish who said what in a crowded room. \oursystem enables a visually-grounded Retrieval-Augmented Generation (RAG) pipeline. Instead of discarding the separated audio buffers, the system can transcribe them into speaker-tagged text segments. These segments are embedded and stored in a vector database. A user can then query an LLM: \dq{What did Peter say about the deadline?}. The LLM retrieves the specific speech segment associated with Peter's visual ID. This application relies on the system's ability to maintain a persistent identity for sound sources over time. As noted in Sec.~\ref{sec:scalability}, future iterations could leverage stronger visual identification embeddings to ensure that the Speaker tag remains consistent throughout long-duration interactions.}



\section{Technical Evaluation}
\label{sec:tech_eval} 

We performed a technical evaluation to assess the performance of \oursystem. Our goal was to quantitatively measure the audio separation quality of our system and validate the separation performance of our distilled real-time models. To do this, we compared \oursystem against several state-of-the-art baselines using objective metrics for both speech and music separation tasks. \change{A supplemental evaluation on the trade-off between model performance and interactive latency is provided in Sec. ~\ref{sec:real-time-analysis}.} This section details our experimental setup and presents an analysis of the results.

\change{
\begin{table*}[t]
\centering
\caption{\change{Comparison of MoXaRt against baselines and audio-visual sound separation benchmarks.}}
\label{tab:system_comparison}
\resizebox{\textwidth}{!}{%
\begin{tabular}{lccccc}
\toprule
\textbf{Reference System} & \textbf{Architecture} & \textbf{Target Domain} & \textbf{Object Selection} & \textbf{Max Sources} & \textbf{Real-Time} \\
\midrule
Sound of Pixels & End-to-End & Music Only & \cmark & 2 & \textbf{\cmark} \\
DAVIS & End-to-End & Universal\textsuperscript{\textdagger} & \xmark & 3 & \xmark \\
AudioScopeV2 & Cascaded & Universal & \xmark & 4 & \cmark \\
AV-MossFormer2 & End-to-End & Speech Only & \xmark & 3 & \xmark \\
\midrule
\textbf{MoXaRt (Offline)} & \textbf{Cascaded} & \textbf{Universal} & \textbf{\cmark} & \textbf{8} & \textbf{\xmark} \\
\textbf{MoXaRt (Real-Time)} & \textbf{Cascaded} & \textbf{Universal} & \textbf{\cmark} & \textbf{8} & \textbf{\cmark} \\
\bottomrule
\multicolumn{6}{p{0.9\linewidth}}{\change{\footnotesize \emph{Note: Universal denotes models capable of separating both speech and music sources. Universal\textsuperscript{\textdagger} indicates the architecture is domain-agnostic , but requires separate training for specific domains (speech vs. music).}}} \\
\end{tabular}%
}
\end{table*}
}

\subsection{Methods}
\subsubsection{Dataset}
\label{sec:tech_eval_dataset}

To evaluate \oursystem in realistic and challenging acoustic environments, we collected a new audio-visual dataset, as existing corpora do not adequately capture the complex, mixed-source scenarios relevant to interactive XR applications. Our dataset consists of 30 unique, one-minute-long recordings categorized into three main types: speech-only, music-only, and mixed-source.

The speech-only subset contains 15 recordings designed to test performance on multi-speaker conversations of increasing complexity. These scenarios feature one (C1), two (C2), three (C3), four (C4), and five (C5) concurrent speakers, with three unique recordings for each condition. The speaker configurations simulate realistic social gatherings, ranging from individuals speaking to multiple simultaneous conversations. For instance, the five-speaker condition features two concurrent conversational pairs and one additional independent speaker. The music-only subset comprises 9 recordings of instrumental performances featuring guitar, flute, piano, and violin. These include solo performances (C6), various duet combinations (C7), and trios (C8). Finally, the mixed-source subset contains 6 recordings that combine instrumental music and speech, featuring scenarios with one speaker accompanied by two instruments (C9) and two speakers accompanied by a single instrument (C10). 

We recorded this dataset from a user's egocentric viewpoint. Our setup (Fig.~\ref{fig:recording_setup}) featured a Meta Quest 3 headset on a stand, outfitted with a Rode Wireless GO II wireless system to capture mono audio input. A tripod-mounted iPhone recorded a corresponding third-person video of each scene. Further details on the recording environment and procedure are located in Appendix~\ref{appendix:tech_eval_dc}.

\begin{figure}[h]
\centering
\includegraphics[width=\columnwidth]{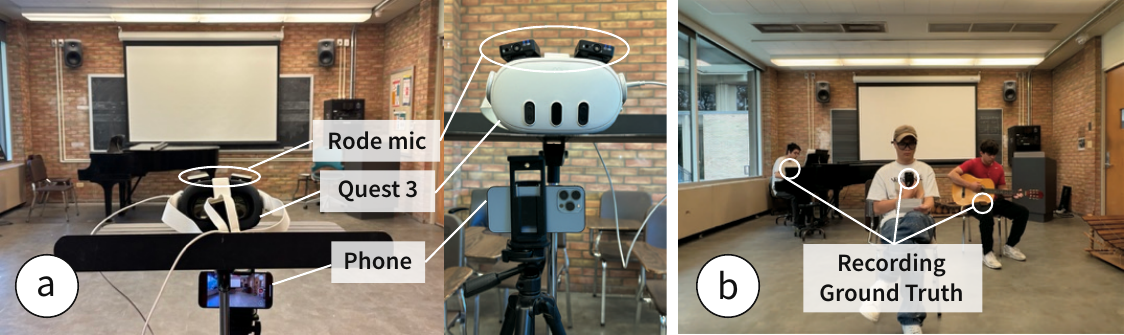}
\caption{Data collection setup. (a) A Meta Quest 3 with two attached Rode Wireless GO II microphones captures the input mono-audio, while a stationary iPhone records a third-person view of the scene. (b) The three sound sources (two instruments and one speaker), each close-mic'd to record their individual ground-truth audio signal.}
\label{fig:recording_setup}
\end{figure}

\subsubsection{Baselines}
To contextualize \oursystem's performance, we compared it against a spectrum of baselines representing diverse and challenging architectural paradigms. \change{Table~\ref{tab:system_comparison} provides a high-level comparison of our system's architectural capabilities against these baselines.} We selected the following systems for comparison:
\begin{itemize}
    \item Raw Audio: We use the original, unprocessed audio to establish a lower performance bound for all metrics.
    \item \change{Sound of Pixels \cite{SoundOfPixels2018}: An end-to-end pixel-level analysis method for filtering specific sound sources, primarily designed for musical instruments.}
    \item AudioScopeV2 \cite{audioscopev2}: An on-screen sound separation system that employs cross-modal and self-attention networks to capture fine-grained audio-visual dependencies. Its inclusion is particularly relevant as our online student models are derived from its architecture.
    \item AV-Mossformer2 \cite{mossformer2}: We use the AV MossFormer2 TSE 16K model from the ClearerVoice-Studio project \cite{clearervoice}. This system serves as a strong point of comparison for contemporary audio-visual methods.
    \item DAVIS \cite{DAVIS}: A state-of-the-art generative baseline for audio-visual separation that employs diffusion and flow matching. To ensure a fair comparison, we trained a 3-source flow matching model (num\_mix=3) using the official recipe. See Appendix~\ref{appendix:davis} for training details.
\end{itemize}

\subsubsection{Metrics}
To quantitatively evaluate the performance of \oursystem and the baselines, we use two widely-adopted sets of metrics:
\begin{itemize}
    \item Word Error Rate (WER): To measure speech intelligibility, we use Gemini 2.5 Pro to transcribe the separated speech audio and calculate the WER against the ground truth transcript.
    \item Deep Noise Suppression Mean Opinion Score (DNSMOS): To assess perceptual audio quality, we use the DNSMOS \cite{dnsmos} metric (scale 1--5). We report the three standard sub-metrics: SIG (signal clarity), BAK (background suppression), and OVRL (overall quality).
\end{itemize}

\subsubsection{Results and Discussion}
\label{sec:tech_eval_results}

The results of our objective evaluation on separation accuracy and quality are presented in Table~\ref{tab:tech_eval_results}. The main takeaways are as follows:

\begin{table*}[h] 
    \centering
    \caption{Objective evaluation of separation accuracy and quality.}
    \label{tab:tech_eval_results}
    \begin{tabular}{lcccc}
        \toprule
        Method   & WER\textsuperscript{\textdagger} $\downarrow$ & DNSMOS-SIG $\uparrow$ & DNSMOS-BAK $\uparrow$ & DNSMOS-OVRL  $\uparrow$ \\ 
        \midrule
        Raw Audio   & 0.6297 ($\pm$ 0.4371) & 3.7740 ($\pm$ 0.1977) & 2.4796 ($\pm$ 0.3928) & 2.8944 ($\pm$ 0.1673) \\ 
        \midrule 
        Sound of Pixels & 0.8636 ($\pm$ 0.1787) & 3.3212 ($\pm$ 0.2467) & 2.9535 ($\pm$ 0.2935) & 2.7291 ($\pm$ 0.2004) \\
        \midrule
        DAVIS & 0.7259 ($\pm$ 0.2484) & 3.3212 ($\pm$ 0.2467) & 2.9535 ($\pm$ 0.2935) & 2.7291 ($\pm$ 0.2004) \\
        \midrule 
        AudioScopeV2  & 0.5263 ($\pm$ 0.3358) & 3.5027 ($\pm$ 0.5029) & 3.0236 ($\pm$ 0.4497) & 2.9349 ($\pm$ 0.3074) \\
        \midrule 
        AV-Mossformer2\textsuperscript{\textdagger} & 0.3956 ($\pm$ 0.2847) & 3.7245 ($\pm$ 0.2813) & 3.2564 ($\pm$ 0.4728) & 3.0759 ($\pm$ 0.2892) \\
        \midrule 
        \textbf{\oursystem Offline}& \textbf{0.3824 ($\pm$ 0.2775)}& \textbf{3.6625 ($\pm$ 0.3059)}& \textbf{3.019 ($\pm$ 0.6078)} & \textbf{2.9949 ($\pm$ 0.2585)} \\
        \midrule 
        \textbf{\oursystem Real-Time}& \textbf{0.4990 ($\pm$ 0.3156)}& \textbf{3.5015 ($\pm$ 0.4976)}& \textbf{3.0065 ($\pm$ 0.4266)} & \textbf{2.9364 ($\pm$ 0.3060)} \\
        \bottomrule
    \end{tabular}
    \vspace{5pt} 
    \begin{center}
        \footnotesize \textsuperscript{\textdagger} Calculations are done over the sessions that contain human speech.
    \end{center}
\end{table*}

\begin{enumerate}
    \item \oursystem Offline Achieves Superior Intelligibility. Our offline model, \textbf{\oursystem Offline}, achieves the best performance in terms of speech intelligibility, securing the lowest \textbf{Word Error Rate (WER)} of \textbf{0.3824}. This result surpasses all baselines, including the highly competitive AV-Mossformer2.
    \item Trade-offs in Perceptual Quality. While \oursystem excels in intelligibility, \textbf{AV-Mossformer2} obtains higher scores on the \textbf{DNSMOS} metrics. This suggests it produces audio with slightly better perceived signal clarity and background suppression, potentially at the cost of some word-level accuracy.
    \item Real-Time Model Validates Distillation. The \textbf{\oursystem Real-Time} model demonstrates the effectiveness of our knowledge distillation approach. It significantly outperforms its architectural basis, \textbf{AudioScopeV2}, achieving a much lower \textbf{WER (0.4990 vs. 0.5263)} while maintaining nearly identical DNSMOS scores.
    \item Performance of Other Baselines. The generative baseline, \textbf{DAVIS}, performed poorly on our dataset with its default configuration, likely due to sensitivity to acoustic characteristics without specific hyperparameter tuning. \textbf{Sound of Pixels} similarly struggled with the complex mixtures in our dataset, resulting in higher WER compared to our cascaded approach.
\end{enumerate}

\section{User Study}
\label{sec:user_study}
We evaluated \oursystem's performance and user experience through a within-subjects study (N=22) across six challenging XR scenarios. Our mixed-methods approach contains both objective performance on listening comprehension tasks and subjective 1-10 scale ratings of key experiential factors.

\begin{figure*}[ht]
\includegraphics[width=2\columnwidth]{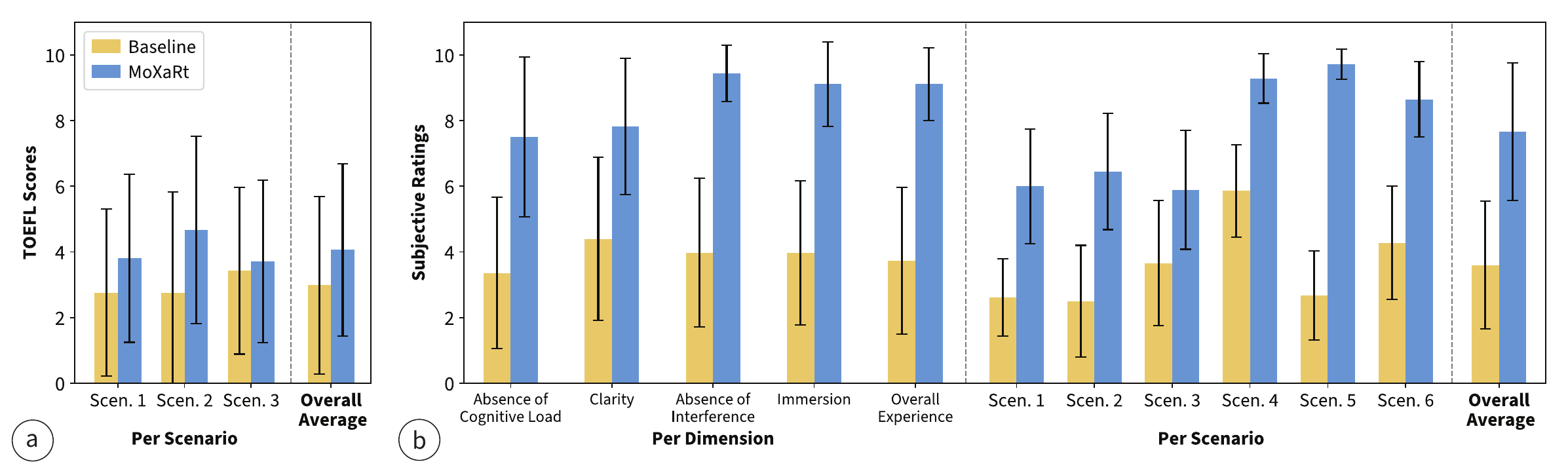}
\caption{\change{User study results. (a) Average Listening Comprehension Performance per scenario and the overall average. \oursystem led to a statistically significant 36.2\% improvement in scores compared to the baseline (p = 0.0058). (b) Average subjective ratings (on a 10-point scale, higher scores indicate better experience) per dimension for the five core experiential dimensions and per scenario and the overall average. Participants rated their experience significantly higher with \oursystem compared to the baseline (all p < 0.001).} Error bars on the average represent the standard deviation.}
\label{fig:user_study_qual}
\end{figure*}

\subsection{Participants}

We recruited 22 participants (8 female, 14 male; mean age 23.2, STD = 2.72). The group included 5 native and 17 non-native English speakers to provide a diverse cohort for the communication tasks. All participants reported normal hearing capabilities and were compensated \$30 for their time.

\subsection{Apparatus and Environment}

The study was conducted using a Meta Quest 3 headset. To create dynamic and realistic yet controlled environments, scenarios involving conversations and musical performances featured pre-recorded actors and professional musicians via Quest's see-through camera in panoramic view. The actors performed standardized TOEFL scripts, while the instrumentalists (a piano, violin, and cello trio) performed both synchronized and unsynchronized pieces. The professional musicians and actors involved in the setup were compensated \$100 for their contribution. \change{We chose to record these scenarios and process them using a consistent real-time model to control for variance in speech and musical performance across participants.}

\subsection{Tasks and Scenarios}

\change{Each participant completed all six scenarios both with and without MoXaRt (baseline), presented in a counterbalanced order.}

\subsubsection{Task 1: Listening Comprehension}
 Participants were instructed to focus on a target conversation and subsequently answer TOEFL comprehension questions. The task was conducted across three distinct scenarios, with one easy, one medium and one hard in difficulty level \change{from Educational Testing Service (ETS) TOEFL Practice Online (TPO) tests\footnote{https://www.ets.org/toefl.html}}. \change{Two conversations of equal difficulty were used for each scenario, and both the order of conditions (MoXaRt and Baseline) and the pairing of conversations with conditions were counterbalanced across participants}:
\begin{itemize}
\item \textbf{Scenario 1: Social Dining with Conversation over Music.} Focusing on a companion’s voice over loud background music from a piano bar. \change{The conversations lasted 2 minutes 50 seconds and 3 minutes respectively.}
\item \textbf{Scenario 2: Multi-Speaker Conference with Distant Male Voices.} Focusing on a more distant conversation between male speakers while a competing conversation occurred nearby. \change{Each conversation lasted 2 minutes 45 seconds.}
\item \textbf{Scenario 3: Multi-Speaker Conference with Distant Female Voices.} Focusing on a more distant conversation between female speakers while a competing conversation occurred nearby. \change{Each conversation lasted 3 minutes 30 seconds.}
\end{itemize}

\subsubsection{Task 2: Interactive Audio Remixing}
Participants could freely manipulate the volume of individual sound sources in real-time. This task was conducted in three scenarios:
\begin{itemize}
\item \textbf{Scenario 4: Orchestra Evaluation.} Adopting a music teacher's perspective to isolate and evaluate one student's performance within a synchronized trio. \change{Each performance lasted 3 minutes 10 seconds.}
\item \textbf{Scenario 5: Rehearsal Evaluation.} Adopting the same teacher perspective, but in a chaotic environment where the trio's members practiced unsynchronized pieces. \change{Each performance lasted 3 minutes 9 seconds.}
\item \textbf{Scenario 6: Concert Interference.} Adopting an audience member's perspective to filter out distracting chatter from nearby people and focus on the musical performance. \change{The performances lasted 2 minutes 53 seconds and 3 minutes 1 second respectively.}
\end{itemize}
\change{For scenarios 4 and 5, the specific instruments participants evaluated were pre-assigned and counterbalanced.}

\subsection{Procedure}
Upon arrival, participants were briefed on the study's purpose and gave informed consent. They were then familiarized with the XR equipment and the interaction mechanics of \oursystem during a 5-minute tutorial session. Before the main tasks, each participant completed a practice trial to ensure they understood the objectives.

Each scenario was completed under both the \oursystem Enabled  and Baseline conditions. To counterbalance the condition order, the conversation-to-condition pairings, and the target instrument designations (piano, violin, or cello), the study was designed for multiples of 12 participants ($2 \times 2 \times 3$). Although 24 participants were originally recruited, two were unable to attend. As a result, the first 12 participants were fully counterbalanced, while the remaining 10 followed a randomized order.

Prior to each scenario, participants were given 30 seconds to preview the questionnaire. This allowed them to familiarize themselves with the specific conversation details to monitor, the target musical instrument for evaluation, and the criteria for subjective rating. After each condition within a scenario, they removed the headset to answer the relevant TOEFL questions within 3 minutes and complete the subjective rating questionnaire within 1 minute. A short break was provided between each scenario. The entire study session lasted approximately 60 minutes.


\subsection{Measures}

We collected both objective and subjective data to evaluate the system's performance and the user's experience.

\subsubsection{Objective Measures}
Our primary objective metric was Listening Comprehension Performance. This was measured by the participant's score on the TOEFL questions following each scenario, with 2 points awarded for each correct answer. A higher score indicates a greater ability to successfully comprehend and retain information amidst distractions.

\subsubsection{Subjective Measures}
After each scenario, participants provided subjective ratings on a 1-10 scale to quantify their experience. These ratings captured five core dimensions: \change{Absence of Cognitive Load}, Clarity, \change{Absence of Interference}, Immersion, and Overall Experience. 


\subsection{Results}
\subsubsection{Objective Results}
To objectively evaluate the impact of \oursystem on a user's ability to understand speech in noisy environments, we analyzed the Listening Comprehension Performance scores from Scenarios 1, 2, and 3. 
\change{
For Scenario 1, participants achieved a 38\% higher score with \oursystem (M=3.81, SD=2.56) compared to the baseline condition (M=2.76, SD=2.55). For Scenario 2, participants achieved a 69\% higher score with \oursystem (M=4.67, SD=2.85) compared to the baseline condition (M=2.76, SD=3.06). For Scenario 3, participants achieved a 8\% higher score with \oursystem (M=3.71, SD=2.47) compared to the baseline condition (M=3.43, SD=2.54).
}
\change{A two-way RM ANOVA was used to analyze the performance scores and Aligned Rank Transform~\cite{art-chi11} was applied before ANOVA to correct non-normality. ANOVA revealed that there were significant effects of \textit{condition} (F = 11.955, p = 0.00248) to the performance, while \textit{scenario} had no significant effect (F = 0.402, p = 0.672). There was no significant \textit{condition} $\times$ \textit{scenario} interaction effect of performance (F = 1.343, p = 0.272). }

In addition, participants on average achieved a 36.2\% higher total score (out of 10 points) with \oursystem (M=4.06, SD=2.06) compared to the baseline condition (M=2.98, SD=2.05). This improvement in performance was statistically significant (p = 0.0058), underscoring the substantial and reliable benefit of using the system. Notice that 1 out of 22 participants (P9) is excluded from the objective result analysis, since their result is >2.5 SDs outside of the scoring distribution of all participants, therefore an outlier.

These results are summarized in Fig. \ref{fig:user_study_qual} (a). The data reveals a consistent trend where participants achieved significantly higher scores with \oursystem enabled compared to the baseline. Across all three tested scenarios, our system yielded superior performance metrics. Notably, the most substantial improvements were observed in participants who performed poorly in the baseline condition; these individuals showed drastic performance gains when using our system, demonstrating its effectiveness in acoustic environments that users find particularly challenging.

\subsubsection{Subjective Results}

To evaluate user experience, we analyzed 1-10 scale ratings from the interactive audio tasks. Paired-samples t-tests were conducted to compare the \oursystem Enabled and Baseline conditions across five core dimensions. Participants rated their experience significantly higher with \oursystem across every dimension ($p < 0.001$). Specifically, they reported a significantly greater absence of cognitive load ($M=7.50, SD=2.44$ vs. $M=3.36, SD=2.31$), higher Clarity ($M=7.82, SD=2.08$ vs. $M=4.40, SD=2.49$), reduced interference ($M=9.44, SD=0.86$ vs. $M=3.98, SD=2.27$), and greater immersion ($M=9.11, SD=1.28$ vs. $M=3.97, SD=2.19$). Reflecting these benefits, the Overall Experience with \oursystem ($M=9.11, SD=1.11$) was rated significantly higher than the baseline ($M=3.73, SD=2.24$).

We further analyzed subjective ratings across individual scenarios. Consistent with the dimensional results, participants rated their experience significantly higher with \oursystem in every scenario ($p < 0.001$). Performance gains were substantial across the board: improvements over the baseline ranged from a 58\% increase in Scenario 4 ($M=9.28$ vs. $M=5.86$) to a 264\% increase in Scenario 5 ($M=9.71$ vs. $M=2.67$). Full details for all six scenarios, including Scenarios 1, 2, 3, and 6 (which saw increases of 130\%, 158\%, 61\%, and 102\% respectively), are visualized in Fig. \ref{fig:user_study_qual} (b).

On average, participants provided 113\% higher ratings with \oursystem ($M=7.66, SD=2.10$) compared to the baseline ($M=3.60, SD=1.95$), indicating a robust subjective preference ($p < 0.001$). These findings complement the objective results, demonstrating that performance gains were accompanied by a substantially improved and less mentally demanding user experience.

\section{Discussion and Future Work}
\label{sec:discussion}


Our work demonstrates the feasibility of utilizing real-time, visually-guided source separation as a core primitive for interactive XR audio experiences. While our technical and user evaluations validate the efficacy of the proposed approach, they also reveal specific limitations that provide a roadmap for future research.

\subsection{Model Performance vs. Real-time Analysis}
\label{sec:real-time-analysis}

The overall user experience in \oursystem depends on the balance between two competing factors: audio separation accuracy and interactive latency. Providing models with longer context windows generally yields a superior listening experience with fewer artifacts (\eg lower WER), thereby increasing perceptual clarity at the expense of longer system response times. Conversely, aggressive parallel streaming reduces delay and enhances responsiveness to user actions, but risks degraded separation quality and increased computational overhead. 

In \oursystem, these factors are primarily controlled by the \textit{window\_size} and \textit{chunk\_size} parameters. The \textit{window\_size} determines the duration of the temporal context provided to the model, impacting both separation accuracy and inference time. The \textit{chunk\_size} dictates the frequency of model execution, which governs the computational load and audio playback latency. In our current implementation, we set these parameters to 60 seconds and 1 second, respectively (see Sec. \ref{sec:interface_implementation}). To quantify these trade-offs, we conducted experiments evaluating the effects of \textit{window\_size} and \textit{chunk\_size} on inference time, WER, computational burden, and total interactive latency.

\begin{figure}[h]
\includegraphics[width=\columnwidth]{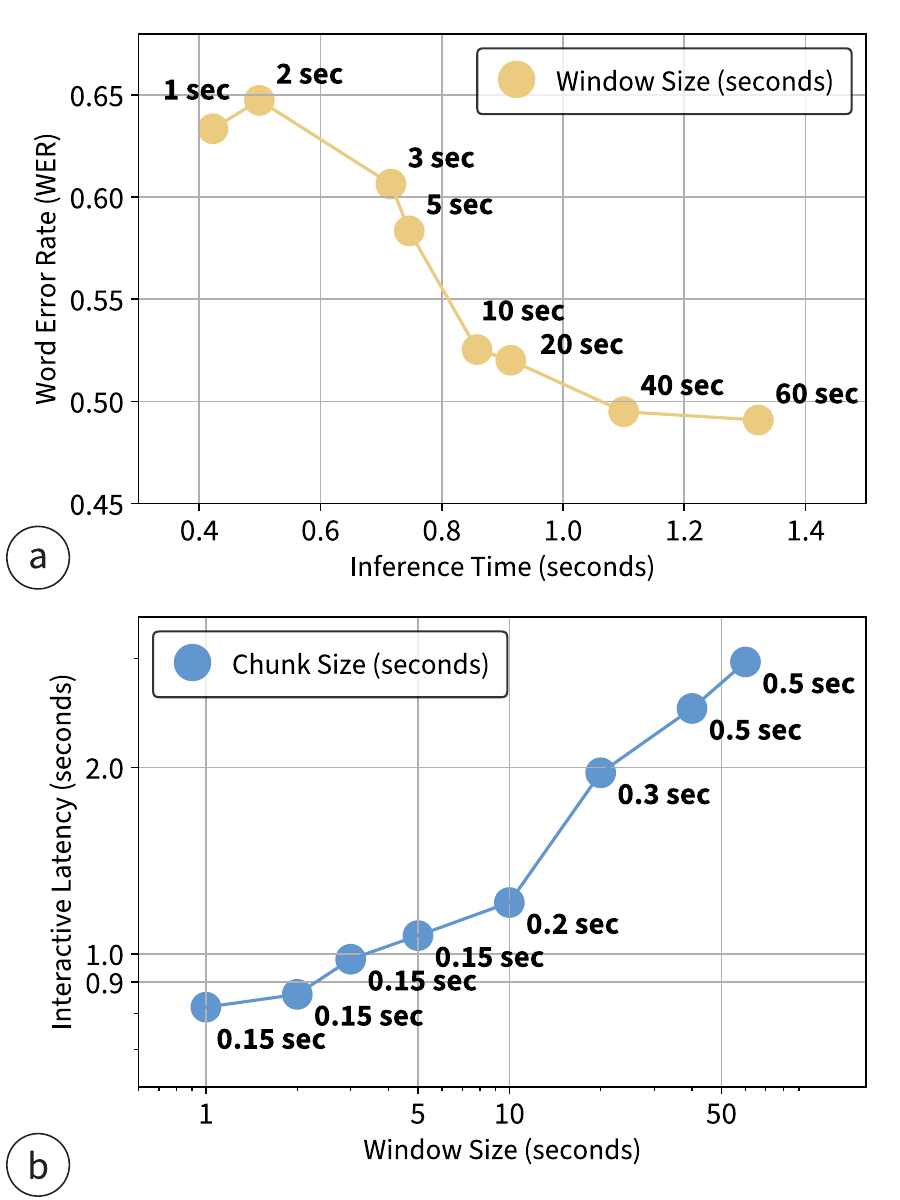}
\caption{Model Performance vs. Real-time Analysis. (a) The size of the context window for model input (\textit{window\_size}) determines the trade-off between WER and inference time. (b) Reducing \textit{chunk\_size} can optimize interactive latency to as low as 0.82s, at the expense of audio separation quality. }
\label{fig:real_time_analysis}
\end{figure}

To examine the trade-off between audio separation accuracy and interactive latency, we measured WER and model inference time across a range of window sizes on a log-scale up to 60 seconds. We processed our dataset at each window size and concatenated the results to calculate the corresponding WER. As illustrated in Fig.~\ref{fig:real_time_analysis} (a), the highest separation accuracy (lowest WER) is achieved with the largest \textit{window\_size}, albeit at the cost of the longest inference time.

To evaluate total system delay, we measured the interactive latency using the smallest operational \textit{chunk\_size} for each given \textit{window\_size}. Although \textit{chunk\_size} can theoretically be reduced indefinitely, it is lower-bounded by the computational overhead required to maintain continuous streaming. We define interactive latency as the delay between a visual event and its corresponding audio rendering through \oursystem, averaged over three measurements across three experimental runs. As shown in Fig.~\ref{fig:real_time_analysis} (b), interactive latency decreases alongside \textit{window\_size}. By utilizing a 0.15-second chunk size with a 1-second window, we achieved an interactive latency of 0.82 seconds. These results suggest that chunk and window sizes can be jointly optimized to balance system responsiveness with separation quality. Further investigation is required to determine how such delays are perceived by users and their ultimate impact on the subjective experience.

\subsection{Deployment Constraints} 
While our results show a promising path towards interactive audio-visual XR, the current implementation faces practical deployment constraints. First, the reliance on visual cues means separation degrades if the source is fully occluded or leaves the camera's field of view. Future iterations could mitigate this by implementing audio-only fallback models when visual data is unavailable. Second, the computational cost of running concurrent detection, tracking, and separation models currently necessitates a tethered PC setup. Migrating \oursystem to standalone headsets will require significant optimization, such as model quantization and on-device NPU acceleration. Another possible solution is to match the video see-through streaming to the processed audio with the same amount of constant latency; however, the effectiveness of such an approach regarding motion-to-photon latency requires further study.

\subsection{Scalability and Persistence}
\label{sec:scalability}
A key challenge in deploying \oursystem to the wild is handling the scale and dynamism of real-world acoustic environments. While our cascaded architecture supports dynamic source numbers, performance naturally degrades as the number of concurrent speakers increases beyond four \cite{DAVIS}. We argue that the solution is not simply larger models, but reframing the problem as attention-driven separation \cite{shin2025saliency}. Instead of attempting to separate all $N$ sources simultaneously, future work should leverage gaze-based prioritization to isolate only the $K$ sources relevant to the user, transforming an intractable signal processing task into a manageable HCI problem.

Furthermore, the system must maintain temporal consistency in these dynamic scenes. Our current window-based processing can be susceptible to \dq{track swapping,} where a speaker who is temporarily silent or occluded is assigned to a different audio channel upon reappearing. Future work will address this by integrating robust identification embeddings (\eg \textit{ArcFace}\cite{ArcFace}) to anchor audio streams to specific individuals, ensuring that a user's volume adjustments persist throughout long-duration, multi-turn interactions.

\subsection{Baseline and Interaction Design}


In our human evaluation, we focused on comparing user experiences with and without \oursystem. We omitted additional baselines because, to our knowledge, no existing system supports real-time, interactive control over a user's authentic soundscape. As discussed in Sec. ~\ref{sec:tech_eval}, state-of-the-art separation models are typically designed for offline processing or restricted to specific domains, such as speech-only, lacking the throughput necessary for live interaction. Similarly, commercial noise-canceling headsets only filter audio by broad categories (\eg stationary noise versus speech) rather than isolating individual sound sources. These devices do not support the granular remixing enabled by our system and are incapable of the core interaction in our study: the selective amplification of a specific target within a multi-source environment. Consequently, they were excluded as comparable baselines.

However, we recognize that our current interaction design, which relies on handheld controllers, represents a limitation for social XR contexts. Future iterations of \oursystem could benefit from more natural input modalities to provide a more seamless user experience. Potential avenues for future research include gaze-dwelling for source selection, micro-gestures for precise volume control, or neural interfaces (EMG/EEG) for discrete, hands-free interaction \cite{shin2025context, zhu2025eit}.

\change{
\subsection{Ethical and Social Implications}
Beyond technical hurdles mentioned in the previous sections, the capacity to selectively filter the acoustic environment introduces complex social and ethical dynamics. Unlike noise-canceling headphones, which signal global isolation, AR-based selective filtering is invisible to bystanders. This raises questions regarding consent and shared reality: does a user have the ethical right to tune out a specific person in a collaborative setting, or silence a specific instrument in a live orchestra, without the source's knowledge or consent? Such capabilities fundamentally alter the social contract of presence. As these technologies mature, we must consider design frameworks that preserve social transparency, perhaps through external signaling (\eg a busy indicator on the headset) or protocols that balance user agency with bystander awareness.}

\section{Conclusion}

In this paper, we presented \oursystem, the first XR system to integrate real-time, audio-visual sound separation as a core primitive for direct, interactive control over a user's authentic soundscape. Our cascaded audio-visual transformer model leverages visual anchors (faces and instruments) to perform robust, real-time separation of individual sources from a single-channel input. Our comprehensive evaluation, performed on a new, complex audio-visual dataset we contributed for this task, confirmed our model's state-of-the-art technical performance. Furthermore, our 22-participant user study confirmed that \oursystem significantly improves communication clarity, boosting listening comprehension by 36.2\% ($p=0.0058$), reduces cognitive load, and enhances the overall user experience with a new sense of interactive sound control.

\balance
\bibliographystyle{ACM-Reference-Format}
\bibliography{sections/references}

\appendix
\newpage
\onecolumn
\section{Technical Evaluation Dataset Collection Setup}
\label{appendix:tech_eval_dc}
\subsection{Recording Environments}
We recorded our datasets in the following two acoustic environments:

\begin{enumerate}
    \item A small study room (approx. 120 sqft) was used for scenarios C1--C5, C10, and C6-1.
    \item A large music school classroom containing a piano (approx. 1500 sqft) was used for scenarios C7--C9, C6-2, and C6-3.
\end{enumerate}

\subsection{Recording Hardware}
\begin{figure}[h]
    \centering
    \includegraphics[width=0.6\linewidth]{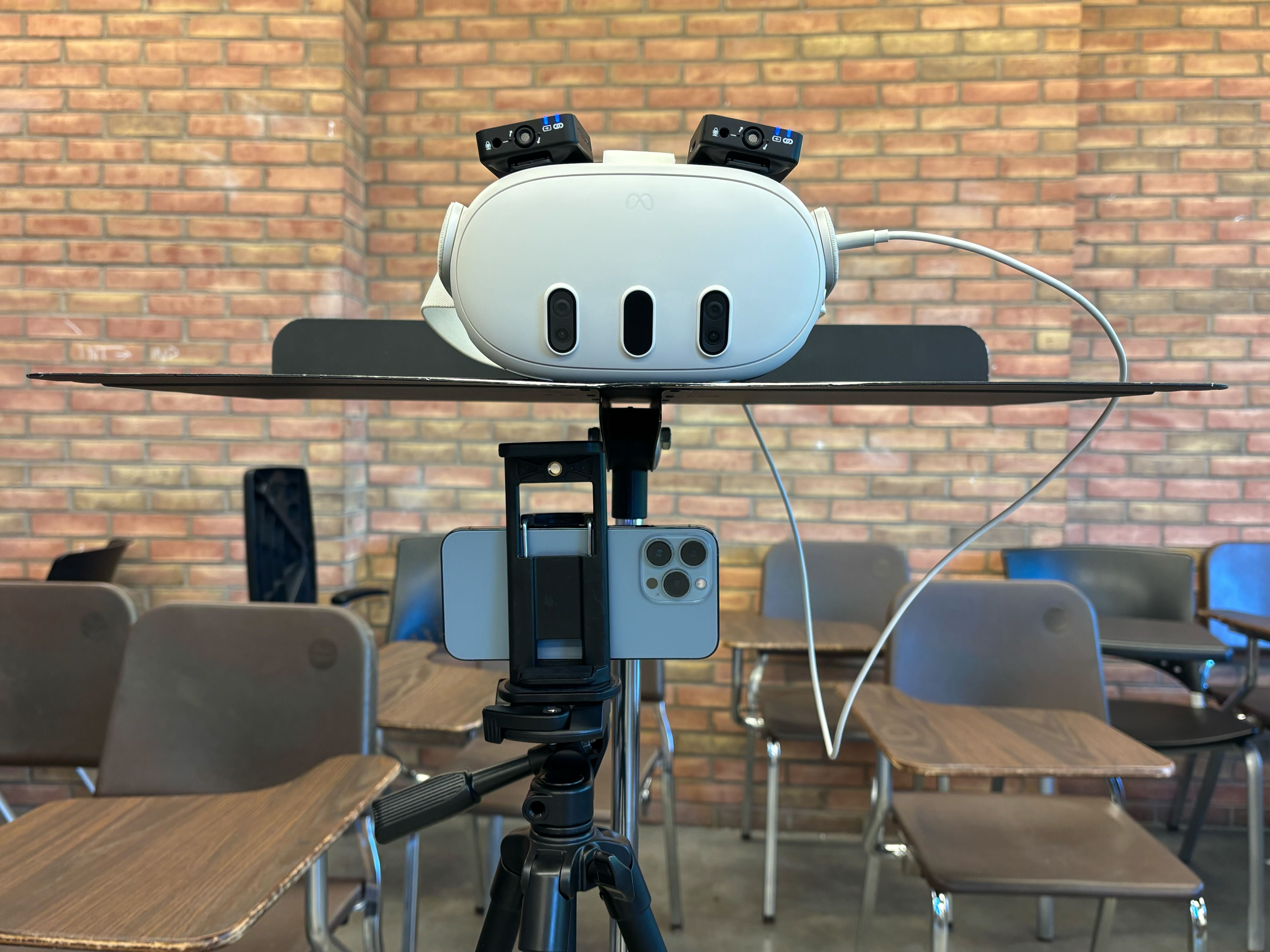} 
    \caption{Data collection setup: A Meta Quest 3 with 2 attached Rode Wireless GO II microphones captures the mono audio, while a stationary iPhone records a third-person view of the scene.}
    \label{fig:recording_setup}
\end{figure}
We employed a dual-system setup for data capture, as depicted in Figure~\ref{fig:recording_setup}. The primary system was designed to capture a stable, user-centric perspective using a Meta Quest 3 headset mounted on a stand at approximately head-level. To record the primary mono audio that serves as the input to our system, a Rode Wireless GO II dual-channel microphone system was attached to the headset, and set to mono-mode to mimic the auditory input via microphones on head mounted devices. Simultaneously, a second system—a stationary iPhone (iPhone 13 Pro, or an iPhone 15 Pro for scenarios C8 and C9) on a tripod—recorded a third-person view of the entire scene for video reference.

\subsection{Speech Content Generation}
The spoken content for our recordings was sourced to reflect realistic use cases.
Single-person speech segments were transcribed from CNN Student News broadcasts. Two-speaker dialogues were generated using ChatGPT (gpt-4o) with prompts such as, \dq{Could you help me generate some simulated TOEFL-style conversations between two people, each about one minute long?} The generated content spanned daily conversations, academic discussions, and professor-student interactions.

\subsection{Data Synchronization}
To ensure precise alignment between the centrally recorded video stream and the multiple independent ground truth audio streams, we employed a manual synchronization method. At the start of each take, a single, sharp clap was performed in the scene. In post-processing, we used Adobe Premiere Pro to align all audio and video tracks by matching the sharp transient peak of the clap event across all recordings. This technique is analogous to the function of a traditional clapperboard in film production.

\section{DAVIS Baseline Training Details}
\label{appendix:davis}

This appendix details the training procedure, evaluation preprocessing, and qualitative performance analysis of the DAVIS baseline model used in our experiments. As our evaluation scenarios involve three concurrent sound sources, we retrained the original DAVIS model, which was designed for two-source separation, to support three-source separation.

\subsection{Model Training}
We trained a 3-source separation model based on the DAVIS framework, specifically using the improved flow-matching variant, DAVIS\_Flow, which demonstrated superior performance in our preliminary tests.

\begin{itemize}
    \item \textbf{Architecture Choice:} We trained both the original DAVIS model and the DAVIS\_Flow variant for 3-source separation on the MUSIC dataset. The DAVIS\_Flow model converged to a validation Signal-to-Distortion Ratio (SDR) of approximately 5, whereas the original DAVIS model only reached an SDR of 2. Consequently, we selected the DAVIS\_Flow model for all subsequent experiments and evaluations.

    \item \textbf{Dataset:} We used the MUSIC dataset provided by the official DAVIS repository. The training split remained unchanged. For validation, we generated new 3-source mixtures by first extracting a complete list of non-overlapping audio files from the provided metadata and then randomly sampling and mixing three files from distinct instrument classes, mirroring the logic of the training data loader.

    \item \textbf{Hardware:} The model was trained on a single NVIDIA A40 GPU (40GB VRAM, CUDA 12.8) with 128GB of main memory and 8 CPU cores.

    \item \textbf{Hyperparameters:} We adopted the hyperparameters from the official DAVIS repository with the following necessary modifications for our 3-source, single-GPU setup:
    \begin{itemize}
        \item \texttt{--num\_mix 3} (changed from 2)
        \item \texttt{--num\_gpus 1} and \texttt{--gpu\_ids 0} (changed from a multi-GPU setup)
        \item \texttt{--batch\_size\_per\_gpu 6} (increased from 4 to utilize available memory)
    \end{itemize}

    \item \textbf{Training Time:}
    \begin{itemize}
        \item \textbf{On MUSIC dataset:} Training the 3-source DAVIS\_Flow model from scratch took approximately four days to converge, at a rate of 3--4 epochs per hour.
        \item \textbf{On AVE dataset:} Due to the significantly longer training time on AVE (approx. 1.5 hours per epoch), we fine-tuned the publicly available 2-source pre-trained checkpoint for 3-source separation. The training took about three days, though validation metrics fluctuated without showing significant improvement over the course of training.
    \end{itemize}
\end{itemize}

\subsection{Evaluation Preprocessing}
To prepare our evaluation scenarios for the DAVIS\_Flow baseline, we followed a multi-step preprocessing pipeline.

\begin{enumerate}
    \item \textbf{Segmentation:} We first segmented each one-minute video into smaller, overlapping chunks using the provided shell scripts. For scenarios involving the MUSIC dataset, we used an audio length (\texttt{audLen}) of 66,150 at a sample rate (\texttt{audRate}) of 11,025 Hz, resulting in approximately 6 segments per video. For scenarios involving the AVE dataset, we used an \texttt{audLen} of 110,000 at an \texttt{audRate} of 11,000 Hz, resulting in about 10 segments per video.
    \item \textbf{Model Inference:} We ran the appropriate pre-trained DAVIS\_Flow model on the segmented data corresponding to each evaluation scenario.
    \item \textbf{Concatenation:} The separated audio segments from the model were concatenated to reconstruct the full-length, one-minute audio tracks.
    \item \textbf{Manual Alignment:} Due to the segmentation and concatenation process, minor length mismatches occurred in some outputs. We used Adobe Premiere Pro to manually trim repeated audio artifacts in four of the output files (C5-1, C8-3, C9-1, C9-3) to ensure their length precisely matched the one-minute input.
\end{enumerate}

\subsection{Qualitative Performance Analysis}
Our evaluation revealed clear patterns in the model's performance, highlighting its strengths and weaknesses.

\begin{itemize}
    \item \textbf{Cases of Poor Performance:}
    \begin{itemize}
        \item \textbf{Speaker Separation:} The model generative perform poorly on speaker separation tasks. The output sometimes degrade into white noise, which explains its worse-than-baseline performance on WER, as shown in Table ~\ref{tab:tech_eval_results}. We hypothesize that this is because the AVE dataset, on which the speech separation model was trained, uses broad class labels like \dq{Male speech} or \dq{Female speech} and is not designed for separating multiple instances of the same class (\eg two different male speakers).
        \item \textbf{Piano Separation:} The model also performed poorly when a piano was one of the sound sources. The separated track for the piano was either a distorted copy of another instrument's audio or white noise. This is an expected failure, as the MUSIC dataset used for training does not contain a \sq{piano} instrument class, making this an out-of-distribution task.
    \end{itemize}

    \item \textbf{Cases of Good Performance:}
    \begin{itemize}
        \item The model performed well on two-source musical instrument separation tasks, provided the piano was not one of the sources. It also successfully separated flute and guitar sounds even when accompanied by human speech, demonstrating a degree of robustness in handling mixtures of music and speech.
    \end{itemize}
\end{itemize}

\end{document}